\documentclass[proof]{aastex}
\input epsf.sty
\newcommand{\chr}{$\chi^2_r$}

\shorttitle{BZBJ1058+5628: a new quasi-periodic BL Lac}
\shortauthors{R. Nesci}
\begin{document}

\title{BZBJ1058+5628: a new quasi-periodic BL Lac object \\
from the Asiago plate archive}

\author{R. Nesci }
\affil{Physics Dept., University of Roma La Sapienza, \\
Piazzale A. Moro 2, I-00185 Roma, Italy 
}
\email{roberto.nesci@uniroma1.it}

\begin{abstract}

We present the historic photographic light curves of three little known Blazars (two BL Lacs and one FSRQ), BZB J1058+5628, BZQ J1148+5254 and BZB J1209+4119 spanning a time interval of about 50 years, mostly built using the Asiago plate archive. 

All objects show evident long-term variability, over which short-term variations are superposed.
One source, BZB J1058+5628, showed a marked quasi-periodic variability of 1 mag on time scale of about 6.3 years, making it one of the few BL Lac objects with a quasi- periodic behavior.
\end{abstract}




\section{Introduction}

Long-term ($\geq$ 20 years) optical light curves are still available only for a relatively small number of Active Galactic Nuclei (AGN) and Blazars.
Pioneering work in this field was made by several authors (Pica et al. 1988; Webb et al. 1988) on relatively large samples of sources over a time window of about 20 years. For some sources these curves show long-term trends occurring over time scales of decades, while others show only short-term variability.
  
A few sources had dedicated papers to their long-term optical variability.
Among the best studied sources we recall: OQ~530, whose optical brightness faded at 0.035 mag/yr for about one century (Massaro et al. 2004); S5~0716+71, showing a monotonic brightening trend of 0.11 mag/yr of its mean luminosity over the last 40 years (Nesci et al. 2005); ON~231, with a long term decreasing trend of 0.023 mag/yr followed by an increasing one of 0.07 mag/yr (Massaro et al. 2001); WGA 0447.9-0322, with a long monotonic trend of 0.11 mag/yr, similar to S5~0716+71 (Nesci et al. 2007);  5C 3.178 declining at 0.03 mag/yr over 30 years (Sharov 1995).

In some cases a periodicity of the outbursts has also been found, the best case by far being OJ~287 (Sillampaa et al 1988, Valtonen et al. 2009). Other such sources are AO~0235+164, with a possible 5 years period of strong outbursts (Raiteri et al. 2008 and references therein) and S5~0716+71 (Raiteri et al. 2003) with recurrence of about 3 years. Also in the radio band a few sources have shown periodic outbursts (e.g. 3C 454.3, Ciaramella et al. 2004, Qian et al. 2009). 

The origin of the fast variations in Blazars is generally explained by the relativistic boosting of perturbations moving down a jet pointing close to the line of sight. The relevant quantity for the boosting is the beaming factor $\delta = 1 \times (\Gamma \times (1-\beta~cos\theta)^{-1}$, where $\Gamma$ and $\beta$ are the Lorentz factor and the velocity (in units of the speed of light) of the perturbations' bulk motion and $\theta$ is the angle between the jet and the line of sight.

On the other hand the nature of secular variations is unclear. A suggestive possibility is that they can be associated with changes in the structure and/or direction of the inner jet (see e.g. Kadler et al. 2006). It is difficult, however, to obtain a clear evidence of such changes because it requires long and accurate multifrequency campaigns  with VLBI angular resolution on a sample of several sources: such a study has been done e.g. by Nesci et al. (2005) in the case of S5~0716+71, suggesting that precession of the jet may indeed explain its observed light curve behavior, or Massaro et al. (2004) in the case of OQ~530.

With the successful launch of the Fermi satellite for Gamma ray astronomy, an all-sky monitoring of the Blazars emission has started, which will probably bring to the attention of astronomers a number of poorly known sources. Lists of potential Gamma ray Blazars have been prepared in the framework of the GLAST project by Massaro et al. (2009), containing about 2800 objects, and Sowards-Emmerd et al. (2005), listing about 770 Northern sky sources. For most of these sources very few data exist, basically those allowing their detection and the classification as an AGN. A better knowledge of the properties of these sources will be useful for the interpretation of the Gamma-Ray data now available from the Fermi mission.

To determine the historic light curve of AGNs, an effective way is to use survey plates taken with wide angle instruments, like the Schmidt telescopes, in fields covered over a large time span for patrol of other targets, like Supernovae or variable stars. A good mine of such material is the archive of the Asiago Observatory (http://dipastro.pd.astro.it/asiago/), with its two Schmidt instruments, the 67/92cm, operative between 1965 and 1998, and the 40/50 cm, operative between 1958 and 1992.

We selected therefore, from the Massaro et al. (2009) catalogue, those sources for which no historic optical light curve is still published, nominally bright enough to be well measurable on the plates of the 67/92 cm telescope (B$\le$17.5), without an obvious strong host galaxy around, and for which a large ($\ge$50) number of plates is available spread over a long ($\ge$10 years) time interval, so that a meaningful historic light curve can be derived. Unfortunately, a very small number of sources matched these conditions, mainly due to their optical faintness.

In this paper we report the results of our photometric measurements of these plates and the first optical historic light curve for three sources, BZBJ1058+5611, BZQJ1148+5254 and BZBJ1209+4119.

\section{Photographic data reduction and light curves} 

Plates were digitized at the Asiago Observatory with an EPSON 1680 Plus scanner, a sampling step of 16 micron (1600 dpi) in grayscale/transparency mode and 16 bit resolution.
Plate scanning included also the unexposed borders to measure the plate fog level ($F$). The transformation of the recorded plate transparency $T$ of each pixel into a relative intensity $I$ was obtained applying the simple relation $I=(F-Z)/(T-Z)$, where $Z$ is the instrumental zero level. The best way to determine this level is to use the value of the darkest pixel in the center of the most overexposed star.

Most of the plates were with 103aO emulsion and GG13 filter, similar to the  IIIaJ+GG385 emulsion+filter combination used in the Second Palomar Sky Survey (POSS II): therefore we established, for each field, a photometric comparison sequence selecting about 20 stars and taking their nominal magnitudes from the GSC2.3 catalogue, using the "J" photometric band, which is based on the POSS II.

Instrumental magnitudes for each plate were obtained using IRAF/APPHOT tasks with a fixed aperture of 3.0 pixels for the 67cm Schmidt plates, corresponding to the FWHM of the average stellar profiles. We checked that the selected stars were not variable. The sky value was taken from a concentric annulus of 10 pixels inner radius and 10 pixels width, using the "mode" algorithm for the computation.

A calibration curve for each plate was then obtained using the GSC2.3 values of the comparison stars. These curves were reasonably tight but showed a slight departure from a simple linear relation, so we fitted them with a 2nd order polynomial law. The most discrepant star was then automatically excluded and a new, final fit, recomputed. 

Having a large number of plates for each field, we performed an intercalibration of our reference stars magnitudes. The differences with respect with the GSC2 values were nearly always smaller than 0.1 mag, indicating that this catalogue is a very good starting point to derive historical photographic light curves.
The intercalibrated magnitudes (which we will call B/Asiago) were then used to rebuild the calibration curve for each plate, which were again well fitted by a 2nd order polynomial relation with an appreciably smaller scatter. 

The $B_J$ magnitude of the AGN was then derived from the instrumental one using the calibration curve.
The scatter of the points around the fitting curve depends on the plate quality: typical rms errors are 0.1 mag. These errors were adopted as estimates of our photometric accuracy.

A number of plates were taken without filter, so that a proper systematic correction had to be derived. Indeed an AGN is typically bluer than the average field stars and should appear therefore brighter when the UV part of the spectrum is not cut by the blue filter. 
Unfortunately we did not find plates taken with the two filter/emulsion combinations in the same night (or nearby nights) for our sources. To find the systematic correction between the two combinations, we decided therefore to plot the light curves derived separately from the plates with and without filter and empirically look for a systematic shift between the two.
This trick was successful in the case of 1148+52, for which there was a large overlap for the two light curves: the correction was found to be small, about 0.2 mag, and was adopted for all three sources.

One source (1058+56) was sufficiently bright that could be well detected also on the films taken with the smaller 40/50cm Schmidt telescope. This allowed to extend the light curve back to the years 1962-65. These films were generally unfiltered and had the PANCHRO ROYAL emulsion; very wisely, however, the Observatory policy for the patrol fields taken with the 40/50 Schmidt was to secure sometimes a couple of images taken in the same night with PANCHRO ROYAL and 103aO (unfiltered) emulsion for comparison. We tried to use these couples of films, where the AGN was well exposed, to determine an empirical transformation from PANCHRO ROYAL into 103aO magnitudes, but the intrinsic photometric accuracy of the stars was rather low (0.2 mag) and the number of pairs too small (5) to derive a statistically reliable correction (-0.4 $\pm$0.2 mag). 
Also in this case therefore we overplotted the uncorrected light curves obtained from the 103aO+GG13 plates and from the unfiltered PANCHRO ROYAL films in the years 1965-1968: the systematic shift was derived comparing the average values of the source in the two data sets. The result was found to be in good agreement with that obtained from the 5 plate pairs seen above.
Finally the correction from 103aO (unfiltered) into 103aO+GG13 derived from the 67cm plates was applied to derive the final $B_J$ magnitudes.

For each source a finding chart is given in Fig. 1, 2 and 3, derived from the Asiago plates, where the comparison stars are marked with letters and A is the AGN.
Table 1 to 3 give the star flag (column 1), RA and DEC (J2000) derived from the GSC2.3 (column 2 and 3), adopted B$_J$ magnitude (column 4) and its uncertainty (column 5) derived from the rms scatter in our dataset.

The photometric data of the three AGNs are reported respectively in Tables 5, 6 and 7. We report here just a few lines of Table 5 as example. Column 1 is the plate identification, column 2 is the filter/emulsion, column 3 the JD, column 4 is the rms deviation of the calibration curve, column 5 the B$_J$ magnitude. The meaning of column 2 is: none=103aO unfiltered; PANC= Panchro-Royal emulsion without filter; GG13= 103aO+GG13.

The resulting light curves are shown in Fig. 4, 5 and 6.

Two additional points were obtained from the POSS I and POSS II digitized blue plates available on-line (e.g. at the ESO website), using the same comparison stars used for the Asiago plates and IRAF/apphot for aperture photometry. The two POSS plates have different emulsion/filter combination (respectively 103aO unfiltered and IIIaJ+GG395) so the POSS I magnitude was corrected like the Asiago 103aO unfiltered ones.

Another recent literature point can be derived from the Sloan Digital Sky Survey (SDSS, Collinge et al. 2005), if the source is not saturated, converting the g magnitude into the GSC2.3 B$_J$ scale.
A linear fit on a scatter plot of B$_J$ {\it vs.} g of the comparison stars for each of our fields gave a slope $\sim$1 and a small shift in the zero point. We expected however the presence of a small color term, due to the redder passband of the g filter with respect to the IIIaJ+GG395 combination: given that the AGNs have an (u-g) color definitely bluer than the average field stars, this term should be appreciable.
To this purpose, scatter plots of the (u-g) {\it vs.} (g-$B_J$) magnitudes of the comparison stars for each field were made: the correlation however was not very tight, probably due to the lower accuracy of the GSC2.3 magnitudes, and in any case all the stars were substantially redder than the AGN so that a large extrapolation of the linear fit would be required to evaluate the color correction; eventually we preferred to ignore the color effect (which is in any case about 0.1 - 0.2 magnitudes) and just apply the zero-point correction.

We made furthermore in the Summer 2009 a final photometric point for each source using our Vallinfreda 50cm Newtonian telescope and CCD camera with a Marconi 47-10 back-illuminated chip and standard Johnson-Cousins filters.

All these additional photometric points are also reported in the respective Tables.

\section{Data of Individual Sources} \label{individual}

BZBJ1058+5628
 
The first optical spectrum of this source was published by Marcha et al. (1996) giving a redshift z=0.41 form a strong [O III] emission line. Subsequent spectroscopic observations did not show any more these lines, but only the spectrum of the host galaxy with absorption lines at z=0.144 (Laurent-Muehleisen et al. 1988, Bade et al. 1988). A more recent spectrum can be found in the SDSS (Adelman-McCarthy et al. 2008), which provides the same redshift value (z=0.143) from several absorption lines of the host galaxy and the faint H$\alpha$ + N[II] emission lines. The equivalent width of this blend is about 5 \AA , giving a classification of BL Lac to this source. The host galaxy is well evident in the image of the SDSS, but absolutely invisible in the Asiago plates. To evaluate the galaxy contribution in the B band for our source we used the equivalent width of the G-band absorption line. The catalogue of absorption lines of elliptical galaxies (Trager et al. 1998) gives an average value 5.5 $\pm1.0$ \AA ~for this feature. The SDSS spectrum of our source has a G-band E.W.=1.0 \AA, so that about 80\% of the flux is actually due to the blazar component at 4300 \AA ~rest-frame. Due to the redshift, the observed B band is centered on the CaII H and K lines, where the galaxy contribution is still lower, so that the flux variation amplitude is little affected by the stable underlying galaxy contribution. 

We found 47 plates with 103aO+GG13 and 25 plates without filter taken with the 67/92 Schmidt, covering the time interval 28-11-1965 to 06-01-1992. Their distribution in time however was not well suited to evaluate the systematic difference between the two combinations. Given the brightness of the source, it was also measurable in several (41 PANCHRO ROYAL, 9 103aO) films of the small Schmidt, allowing us to derive its light curve in the years 1962-65. Films taken during the year 1969 were used to determine the systematic difference between the two filter/emulsion combinations.

The Asiago light curve covers about 11000 days and shows a definite monotonic decreasing trend of 2 magnitudes (0.053 mag/year). Superposed on this trend there are wide oscillations with peak-to-peak amplitude of about 1 mag, with a time scale of about 2500 days (7 years).
The Fourier spectrum is dominated by three low frequencies (0.00039, 0.00066 and 0.00082, corresponding to 2550, 1490 and 1220 days) with similar amplitude, but a lot of other frequencies are present with non negligible amplitude.
Making a fit of the light curve keeping only the three low frequencies, the amplitude of the 2550 days period is the largest (0.27 mag) followed by 1220 (0.14) and 1490 (0.11), with a rms deviation from the actual light curve of 0.22 mag.
The deviations are mainly concentrated in some segments of the light curve, and can be modelled only including several much higher frequencies: these deviations can be interpreted as flaring episodes in the jet, overimposed to the basic oscillating behavior of $\sim$ 2500 days. Such random variability is typical of BL Lac objects, so it is expected to be present.
Using only the lowest frequency (which is the strongest) to fit the light curve, the best fit is achieved with a period is of 2590 days and the rms deviation is 0.25, not dramatically larger than the 3-frequencies case. A formal Monte Carlo estimate of the period accuracy, made using the tools of Period04, gives 2590+/-30 days. A sinusoidal curve with period 2590 days and a decreasing trend of 0.053 mag/yr is reported in Fig. 4 to guide the eye.

We measured the source also on the POSS I digitized plate (1953-03-18, 103aO unfiltered, JD 34454) using our comparison sequence: the fit showed a rather large scatter (0.3 mag) and the source was at 16.05, which translate into B$_J$ 16.25 on our scale (see below the discussion of BZQJ1148+52). The same procedure on the POSS II plate (1991-02-09, IIIaJ+GG395, JD 48296) gave a much better fit (rms 0.09) and the source was at $B_J$=16.19, similar to the GSC2.3 catalogue value of 16.04.

The POSS II observation is fully consistent with the Asiago light curve (see Fig. 4), while the POSS I observation is well below the extrapolation of the long term decreasing trend between JD 37,500 and JD 47,000: this is suggestive that this trend may be just part of an oscillating behavior with longer timescale.
Further indications that this is actually the case come from the last points of the Asiago light curve (after JD 47000) and by more recent observations. One is from the SDSS (JD 52263, g=16.50) which transforms into B$_J$=16.20 in the GSC2.3 scale. Another is from our own photometry using the Vallinfreda 50cm telescope  (JD 54995), which gives B=15.33 $\pm$0.16 and therefore tells that the source has returned to the very high state of the early 1960's (see Fig. 4).

\medskip

BZQJ1148+5254

The spectrum of this source in the SDSS shows strong emission lines and a redshift z=1.632: it is therefore classified as a Flat Spectrum Radio Quasar in the Roma BZCat. The image is point-like in the SDSS, as expected for a high-redshift AGN.
We found 67 plates with 103aO+GG13 and 19 plates with 103aO without filter taken with the 67/92 Schmidt, covering the time interval 24-01-1966 to 09-05-1994. The time distribution of the plates for this source is well suited to evaluate the systematic difference between the two dataset: the light curves derived separately were indeed rather similar, with the unfiltered points systematically brighter, as expected. To derived the systematic correction for the unfiltered values we averaged the magnitudes of the AGN in the time interval JD 44000 - 46500, (12 points without filter and 17 with GG13), obtaining $\Delta$M=0.20. This value was adopted also for the other two sources.

The Asiago light curve covers about 9000 days and shows a monotonic decreasing trend of about 1.8 mag (0.073 mag/year) with a slope quite similar to that of 1058+56. The data are scattered around this trend, with no definite shorter time-scales. The photometric accuracy is however lower than in the case of 1058+56, due to the lower flux level of this source. As expected from the visual inspection, the DFT analysis did not show any remarkable frequency in the data besides the seasonal one year time scale of the observations. A couple of points definitely below the average (JD 43311 and 46478) are probably underestimated due to the lower quality of the plates. 

On the POSS I plate (103aO unfiltered, 1950-03-20, JD 33360.85) we measured the source at 15.98 $\pm$0.10, which becomes B$_J$=16.18 in the 103aO+GG13 system.
The POSS I point is therefore brighter than the brightest one in the Asiago light curve, but definitely fainter than the straight extrapolation backwards of the observed trend. 
On the POSS II plate (IIIaJ+GG395, JD 49456.78) we measured the source at B$_J$ 17.20 $\pm$0.07. This observation is about 500 days after the last Asiago photographic point, and is roughly consistent with the general trend.

On the SDSS image (JD 52288) the source was at g=16.93, which becomes B$_J$=17.15 on our scale.  This is suggestive that the monotonic decrease has continued up to the year 2002.

Our CCD photometry from Vallinfreda (JD 54995), however, puts the AGN at B$_J$=16.6, indicating that the source has considerably brightened and is now back at the levels of 1960's.

\medskip

BZBJ1209+4119

The source has an SDSS nearly featureless spectrum, and can be classified therefore as a BL Lac. In the SDSS database it is given a redshift z=0.377 and the image is point-like.

We found 32 useful plates with 103aO+GG13 and 18 plates without filter in the 67/92 Schmidt archive. Most of the plates were centered on the star 2 CVN and some on 67 UMA.
The dataset of two filter/emulsion combinations are not well mixed, so it is not easy to determine from them the systematic effect on the AGN magnitude. We used therefore the correction obtained from 1148+52 as discussed above. The source was too faint to be measurable on the small Schmidt films.

The Asiago light curve covers about 9500 days. The source is rather faint for the 67/92 Schmidt so the photometric accuracy is worse than for the other two sources. In this case a slow brightening trend (0.04 mag/year) seems to be present, with a large dimming episode around JD  45000. The rms deviations of the comparison stars of comparable faintness (S,G,T,Q, see Table 3) from their average values are anyway much smaller than that of the AGN, so that at least in a statistical sense the overall source variation is real. In the final part of the light curve the flux variation of the AGN is smaller, suggesting that part of the variations observed when the source was fainter may be due to its low signal/noise ratio; however, the faint comparison stars were always detected on the Asiago plates, so that we are confident that the low state of the AGN around JD 45000, recorded on several plates, is not just an artifact of photometric uncertainty. 

Due to the low S/N ratio of the data, the low sampling frequency (on average 1 point every six months) and the presence of a single strong dimming feature in the light curve we do not think than a periodicity search is reliable.

On the digitized POSS I plate (1955-03-24, JD 35190) we measured the source at 17.32 $\pm$0.10, which converts into B$_J$=17.52 in our system. On the POSS II plate (1996-03-19, JD 50161) we measured the source at B$_J$=17.57 $\pm$0.04. On the SDSS image (JD 52731) the source is at g=17.98, which becomes B$_J$=18.28 on our scale. We observed the source from Vallinfreda (JD 55040) in the R band, because it was too faint to be detected in B; in this case we used the SDSS r magnitudes for our comparison sequence stars and the source was at about 17.5, very near to the level of the SDSS observation (r=17.62), suggesting that the source did not change very much in the last years.

The slow historic trend shown by the Asiago plates does not extend outside the time window sampled by the Asiago observations. Indeed if we extrapolate the trend to the POSS I and POSS II epochs we would expect 18.1 and 16.8, while the source was at 17.5 in both plates. The two more recent observations of the SDSS and our own found the AGN in a much fainter state, at about 18.3, definitely not on the extrapolated tend (16.6) but rather at a flux level similar to the historic minimum recorded by six Asiago plates. Only a long time monitoring will tell us if such dimmings are a real characteristic of this source.

\section{Discussion}

Pica et al. (1988) and Webb et al. (1988) studied the long-term behavior of several tens of AGNs with photographic plates for 15 to 20 years. For 61 sources they had enough data to morphologically classify their light curve in four types: Class I flickering without long term trend; Class II long term trend with small flickering; Class III long term and short term variability of comparable amplitude; Class IV rare outburst with stable flux level. In the Blazar sample of 22 sources of Webb et al. (1988) there is no significant difference in the frequency of the light curve classes between FSRQ and BL Lac objects, while a marked difference between Blazars and QSO exists in the Pica et al. (1988) sample of 39 sources, which includes a good number of radio steep spectrum Quasars and radio quiet QSO.

To have some physical insight of our sources we report in Table 7 some basic data: column 1 is the name, column 2 the Log(power) at 1.4 GHz, column 3 the absolute R magnitude computed from the USNO B1 catalogue magnitude and literature redshift, column 4 the Radio/Optical flux ratio, column 5 the NIR spectral slope from the JHK magnitudes in the 2MASS catalogue, column 6 the optical spectral slope from the SDSS data. We remark that, at variance with the other two sources, the spectral slopes computed for BZQJ1148+52 have a very poor \chr ~and are therefore marked with ":". Actually the spectrum of this source cannot be well fitted with a power law in neither of the two explored ranges, probably due to the strong emission lines (e.g. C IV equivalent width is 77 \AA) and of the UV bump which falls in the optical due to the source redshift.

It is apparent from this Table that our three sources are flat-spectrum radio-loud objects (Radio/Optical flux ratio $\ge$10) but have different absolute luminosities and show substantially different behaviors in their optical light curve. The strong-lined object is the brightest both in the radio and optical bands. Only BZBJ1058+56 was detected in Gamma-rays by Fermi-LAT (Abdo et al. 2009) and possibly also by EGRET (Bloom et al. 2000).

From the point of view of the overall Spectral Energy Distribution, a much used tool for the classification of Blazars is the $\alpha_{ro}-\alpha_{ox}$ diagram (Padovani and Giommi 1995). On this diagram the Blazars mainly occupy two areas: a horizontal branch and a diagonal branch; a diagonal line of negative slope -1 is a line of constant Radio/X-ray flux ratio. The line at $\alpha_{rx}$=0.75 is the formal border between HBL and LBL sources  (Padovani and Giommi 1995).
Extreme HBL sources are located at the left side of the horizontal branch, extreme LBL at the upper side of the diagonal branch.  For a Synchrotron Self Compton emission model, as the peak of the synchrotron emission of a Blazar moves from lower ($10^{13}$Hz) to higher ($10^{17}$Hz) frequencies the location of the source on this diagram moves from the upper left corner to the lower right one along the diagonal line and then back to the left along the horizontal branch.

We report in Fig. 7 this diagram for the sources in the Roma BZCat, with the positions of our three objects marked.
BZBJ1209+41 and BZQJ1148+52 have $\alpha_{rx}$ larger than 0.75, and are located in the diagonal branch, with BZBJ1209 being the most radio loud. BZBJ1058+56 is already on the horizontal branch and is an HBL, as discussed by Donato et al. (2005) also on the basis of BeppoSAX X-ray spectra. None of them is however an extreme case.

The long-term optical light curves of our three sources are rather different. As a general remark, the overall amplitude variability is anticorrelated with the intrinsic power.

BZBJ1058+56 showed regular oscillations of about 1 mag amplitude, with timescale of $\sim$2300 days, over a monotonic decreasing trend of 0.07 mag/year; it can be classified as Class III and is the source with the larger variability.  Its historic light curve contains 5 outbursts sampled by the Asiago plate archive; this source seems therefore an interesting case of quasi-periodic BL Lac. Further multiwavelength monitoring of this source, should be performed. 

BZQJ1148+52 showed a monotonic decreasing trend with a slope similar to 1058+56 but without the oscillating behavior: the detected short-term variability is comparable to our photometric uncertainty, so no firm conclusions can be derived on their time scale. It can be put into Class II. It showed a substantially smaller variability than the other two BL Lacs, both on short and long time scales.

BZBJ1209+41 showed a slight increasing trend (0.04 mag/year) with large dips in its light curve: it is therefore quite unusual (the opposite of Class I) and deserves further monitoring. Unfortunately it is not bright enough to be easily followed with small telescopes. 

When a time interval of about 50 years in considered, all the long term trends detected in the time window of 27 years sampled by the Asiago plates do not seem to hold, so that they might be considered just as part of longer variability trends.

Which processes can be behind these secular trends? Both physical processes and geometrical effects can be at work: in the first case one can imagine a monotonic variation of the number of radiating electrons, or of the average ambient magnetic field; in the second case, a change in the Doppler boosting factor along our line of sight due to the jet precession. The latter possibility can be considered as an indicator of a massive black hole binary system in the nuclear region (see e.g. Romero et al. 2003).

If we interpret the long term trends of our three sources as due to a slow precession of the jet, as in was supposed to be in the case of S5 0716+71 (Nesci et al. 2005), then the periodicity should be of several 10$^4$ days and therefore comparable to (or even larger than) the human lifetime. 
This poses a strong challenge because observations must be accumulated for several tens of years before any firm conclusion can be reached. A further difficulty for the data interpretations, if the monitoring is not dense enough, could be the occurrence of fast and/or large occasional outbursts/dips, which can mask the long-term trends. 

A strong support to confirm the precession model could come from imaging at high radio frequencies with VLBI techniques, which could detect monotonic variations in the jet direction and/or at lower frequencies showing residuals of radio emission in regions involved by the crossing of the jet in the past (see e.g. Massaro et al. 2004).

Finally, we remark that the detection of the quasi-periodicity of BZB J1058+56 from the Asiago plates suggests that further discoveries could be made using other, still unexplored, photographic plate archives. 



\begin{acknowledgements}
We thank the Asiago Observatory and the Department of Astronomy of Padova University for hospitality for the plate archive search, and Alfredo Segafredo for scanning part of the plates.
The Guide Star Catalogue-II is a joint project of the Space Telescope
Science Institute and the Osservatorio Astronomico di Torino (OATo/INAF). 
This research made use of the CDS (Strasbourg), NED (NASA /IPAC
Extragalactic Database) and SDSS (Sloan Digital Sky Survey) databases.

 
\end{acknowledgements}


\clearpage

\begin{deluxetable}{cccccc}
\tabletypesize{\footnotesize}
\tablecolumns{6}
\tablewidth{0pc}
\tablecaption{Comparison stars in the field of BZB J1058+56 
\label{std1058}}
\tablehead{
\colhead{RA(2000)} & \colhead{DEC(2000)}& \colhead{ident.} & \colhead{$B_{J}$ 
mag} &
\colhead{B/Asiago}  & \colhead{rms}\\
}
\startdata
\hline\\
164.27350&	56.53899&	B&	16.55&	16.62& 0.08\\
164.32579&	56.43882&	C&	17.72&	17.66& 0.08\\
164.34494&	56.44861&	E&	17.68&	17.74& 0.09\\
164.35300&	56.38940&	D&	16.09&	16.00& 0.09\\
164.38134&	56.46707&	F&	17.22&	17.29& 0.09\\
164.47731&	56.45302&	I&	15.61&	15.80& 0.07\\
164.47977&	56.58929&	J&	16.16&	16.38& 0.08\\
164.60784&	56.44601&	K&	16.58&	16.65& 0.08\\
164.61842&	56.54285&	L&	17.01&	16.98&0.09\\
164.62266&	56.43455&	M&	15.58&	15.57& 0.05\\
164.65048&	56.34049&	N&	15.19&	15.33& 0.06\\
164.65070&	56.53979&	O&	15.94&	15.92& 0.06\\
164.67445&	56.50317&	P&	17.73&	17.67& 0.08\\
164.69448&	56.39540&	Q&	15.43&	15.31& 0.06\\
164.75473&	56.33177&	R&	17.12&	17.03& 0.07\\
164.80188&	56.49624&	S&	17.71&	17.60& 0.08\\
164.82489&	56.33905&	T&	15.50&	15.43& 0.09\\
164.84320&	56.37278&	U&	17.18&	17.32& 0.13\\
164.97340&	56.44394&	Y&	16.18&	16.02& 0.16\\
164.97473&	56.48899&	W&	17.43&	17.35& 0.14\\
\enddata
\end{deluxetable}

\begin{deluxetable}{cccccc}
\tabletypesize{\footnotesize}
\tablecolumns{6}
\tablewidth{0pc}
\tablecaption{Comparison stars in the field of BZB J1148+52 
\label{std1058}}
\tablehead{
\colhead{RA(2000)} & \colhead{DEC(2000)}& \colhead{ident.} & \colhead{$B_{J}$ 
mag} &
\colhead{B/Asiago}  & \colhead{rms}\\
}
\startdata
\hline\\
177.21235&	52.93091&	B&	16.14&	16.20& 0.09\\
177.28202&	52.84490&	E&	17.34&	17.27& 0.12\\
177.43974&	52.97246&	F&	17.87&	17.77& 0.14\\
177.43928&	52.98250&	G&	17.77&	17.78& 0.14\\
177.49464&	52.97587&	H&	16.29&	16.22& 0.11\\
177.50878&	52.98667&	I&	17.40&	17.48& 0.15\\
177.44770&	53.04665&	L&	16.67&	16.90& 0.12\\
177.02230&	52.99680&	M&	17.80&	17.83& 0.13\\
177.09700&	53.01454&	N&	17.81&	17.70& 0.13\\
177.10528&	52.90447&	P&	15.66&	15.48& 0.05\\
177.13476&	52.89412&	Q&	15.92&	16.05& 0.13\\
177.19421&	53.00136&	S&	15.98&	15.95& 0.09\\
177.25626&	53.02330&	T&	17.35&	17.27& 0.14\\
176.96426&	52.95703&	U&	15.29&	15.43& 0.08\\
177.15185&	52.84147&	V&	16.94&	16.92& 0.12\\
\enddata
\end{deluxetable}

\begin{deluxetable}{cccccc}
\tabletypesize{\footnotesize}
\tablecolumns{6}
\tablewidth{0pc}
\tablecaption{Comparison stars in the field of BZB J1209+41 
\label{std1058}}
\tablehead{
\colhead{RA(2000)} & \colhead{DEC(2000)}& \colhead{ident.} & \colhead{$B_{J}$ 
mag} &
\colhead{B/Asiago}  & \colhead{rms}\\
}
\startdata
\hline\\
182.48757&   41.35025&   B&  17.02&  17.09& 0.09\\
182.38187&   41.43152&   D&  17.00&  17.06& 0.08\\
182.41075&   41.24259&   E&  16.23&  16.30& 0.07\\
182.23561&   41.42514&   F& 16.98&  17.09& 0.09\\ 
182.39451&   41.35070&   G& 18.34&  18.24& 0.16 \\
182.38566&   41.35420&   H& 16.07&  16.20& 0.12\\
182.19538&   41.40745&    I&  16.59&  16.67& 0.07\\
182.20402&   41.23872&   J&  14.97&  14.97& 0.11 \\
182.33838&   41.27927&  M& 15.36&  15.33& 0.07\\ 
182.29515&   41.21913&  O&  17.49&  17.40& 0.09\\
182.35833&   41.18281&  P&  15.34&  15.29& 0.06 \\
182.29730&   41.24304&  Q&  18.63& 18.42& 0.16\\
182.37593&   41.18086&  R&  17.83& 17.81& 0.13\\
182.37925&   41.37451&  S& 18.39&  18.02& 0.12\\
182.44092&   41.37890&  T& 18.03&  18.39& 0.21\\
182.21330&   41.44316&  W& 18.01& 17.91& 0.16 \\
182.20821&   41.44291&  Z& 17.83&  17.86& 0.12\\

\enddata
\end{deluxetable}

\begin{deluxetable}{cccccc}
\tabletypesize{\footnotesize}
\tablecolumns{6}
\tablewidth{0pc}
\tablecaption{Relevant data for our Blazar sample.
\label{relev}}
\tablehead{
\colhead{Name} & \colhead{log(P$_{1.4}$) } & \colhead{M$_R$}& \colhead{R/Opt} & \colhead{$\alpha _{JHK}$} & \colhead{$\alpha _{ugriz}$} \\
}
\startdata
\hline\\
BZBJ1058+56& 24.9& -25.0& 29.5& -0.58  & -1.35 \\
BZQJ1148+52& 26.8& -28.9& 63.1& -0.16: & -0.97: \\ 
BZBJ1209+41& 25.9& -24.2& 561 & -1.13  & -1.23 \\
\enddata
\end{deluxetable}

\begin{deluxetable}{lcccc}
\tabletypesize{\footnotesize}
\tablewidth{0pc}
\tablecolumns{5}
\tablecaption{Historic $B$ magnitudes of BZB1058+5621 from the Asiago archive 
\label{histBmag}}
\tablehead{
\colhead{Plate} & \colhead{Filter}   & \colhead{JD}    & \colhead{B$_J$} &
\colhead{err}    
}
\startdata
OSC O712 &  none &   34454.7917 &16.25& 0.35\\
 S40 02499&PANC&   37694.2986& 14.92&  0.06 \\
 S40 02529&PANC&   37696.4341& 14.83&  0.08 \\
 S40 02552&PANC&   37698.4230& 15.55&  0.10 \\
 S40 02605&PANC&   37702.3035& 14.73&  0.08 \\
 S40 02615&PANC&   37707.5452& 14.84&  0.09 \\
 S40 02634&PANC&   37719.3348& 15.00&  0.10 \\
 S40 02659&PANC&   37732.3153& 14.88&  0.08 \\
 S40 02691&PANC&   37753.4709& 14.93&  0.08 \\
 S40 02708&PANC&   37760.3709& 14.92&  0.07 \\
 S40 02732&PANC&   37763.3938& 14.83&  0.11 \\
 S40 02755&PANC&   37780.3271& 15.06&  0.09 \\
 S40 02781&PANC&   37786.3313& 14.74&  0.07 \\
 S40 02831&PANC&   37817.4757& 14.67&  0.05 \\
 S40 02845&PANC&   37821.4848& 14.92&  0.21 \\
 S40 02861&PANC&   37824.5278& 15.10&  0.12 \\
 S40 02896&PANC&   37843.4882& 14.74&  0.16 \\
 S40 03291&PANC&   37972.6174& 15.43&  0.09 \\
 S40 03324&PANC&   37992.5924& 15.22&  0.15 \\
 S40 03360&PANC&   37998.5827& 15.54&  0.12 \\
 S40 03400&PANC&   38001.6591& 15.27&  0.10 \\
 S40 03435&PANC&   38005.6243& 15.35&  0.08 \\
 S40 03504&PANC&   38020.5355& 15.47&  0.09 \\
 S40 03600&PANC&   38049.5528& 15.41&  0.09 \\
 S40 03675&PANC&   38055.5382& 15.41&  0.06 \\
 S40 03741&PANC&   38081.4813& 15.34&  0.08 \\
 S40 03817&PANC&   38089.6466& 15.20&  0.10 \\
 S40 03864&PANC&   38106.3757& 15.20&  0.08 \\
 S40 03908&PANC&   38139.3243& 15.35&  0.14 \\
 S40 03946&PANC&   38166.4146& 15.51&  0.10 \\
 S40 04007&PANC&   38193.3931& 15.06&  0.06 \\
 S40 04055&PANC&   38207.5049& 15.40&  0.08 \\
 S40 04138&none&   38316.6216& 15.62&  0.12 \\
 S40 04235&PANC&   38346.6146& 15.40&  0.07 \\
 S90 00060&none&   39093.6000& 15.41&  0.04 \\
 S90 00182&none&   39168.5667& 15.06&  0.04 \\
 S90 00228&none&   39199.4076& 15.64&  0.11 \\
 S90 00473&none&   39475.6201& 15.28&  0.06 \\
 S90 00573&none&   39506.6292& 15.54&  0.10 \\
 S90 00635&none&   39587.3840& 15.35&  0.08 \\
 S90 00651&none&   39609.3285& 15.34&  0.06 \\
 S90 00690&none&   39622.4472& 15.18&  0.04 \\
 S90 01295&none&   39864.5653& 15.48&  0.09 \\
 S90 01466&GG13&   39914.4542& 15.71&  0.14 \\
 S90 01571&GG13&   39942.4750& 15.56&  0.07 \\
 S90 01587&GG13&   39947.3979& 15.68&  0.09 \\
 S90 01623&GG13&   39972.3958& 15.62&  0.06 \\
 S40 06837&PANC&   40150.6035& 15.34&  0.11 \\
 S40 06838&none&   40150.6146& 15.14&  0.34 \\
 S90 02203&GG13&   40260.4757& 15.39&  0.05 \\
 S90 02225&GG13&   40263.4660& 15.30&  0.12 \\
 S90 02230&GG13&   40267.5819& 15.12&  0.08 \\
 S90 02246&GG13&   40271.4701& 15.31&  0.04 \\
 S90 02276&GG13&   40290.3903& 15.81&  0.04 \\
 S40 07158&PANC&   40319.4063& 15.20&  0.13 \\
 S90 02289&GG13&   40319.4861& 15.30&  0.05 \\
 S90 02301&GG13&   40321.5708& 15.33&  0.03 \\
 S90 02315&GG13&   40322.4868& 15.30&  0.04 \\
 S40 07220&PANC&   40327.3757& 15.46&  0.13 \\
 S90 02350&GG13&   40329.5382& 15.51&  0.03 \\
 S40 07301&PANC&   40392.4237& 15.49&  0.11 \\
 S40 07302&PANC&   40392.4333& 15.49&  0.12 \\
 S40 07321&PANC&   40412.4341& 15.60&  0.12 \\
 S40 07494&PANC&   40505.6098& 15.68&  0.13 \\
 S90 04272&none&   40979.4882& 16.34&  0.06 \\
 S90 04212&none&   40999.5375& 16.10&  0.09 \\
 S90 04213&none&   40999.5521& 16.06&  0.06 \\
 S90 05376&none&   41446.3486& 15.82&  0.19 \\
 S90 07799&GG13&   42425.4889& 15.53&  0.05 \\
 S90 07880&GG13&   42459.5257& 15.46&  0.05 \\
 S90 07895&GG13&   42514.3826& 15.59&  0.06 \\
 S90 08415&GG13&   42833.4562& 15.81&  0.10 \\
 S90 08453&GG13&   42863.4681& 15.84&  0.03 \\
 S90 08925&GG13&   43187.3549& 16.36&  0.15 \\
 S90 08949&GG13&   43226.4403& 16.19&  0.10 \\
 S90 09446&GG13&   43540.3833& 16.30&  0.11 \\
 S90 09479&GG13&   43569.3611& 16.49&  0.09 \\
 S90 09818&GG13&   43843.5243& 16.16&  0.09 \\
 S90 09991&GG13&   43966.5528& 16.45&  0.05 \\
 S90 10300&GG13&   44203.5389& 16.38&  0.07 \\
 S90 10318&GG13&   44220.4896& 16.57&  0.04 \\
 S90 10368&none&   44250.3910& 16.67&  0.13 \\
 S90 10393&none&   44273.3826& 16.35&  0.11 \\
 S90 10467&none&   44344.4153& 16.18&  0.07 \\
 S90 10780&none&   44611.5354& 15.73&  0.06 \\
 S90 10816&none&   44630.3521& 15.97&  0.07 \\
 S90 10864&none&   44636.4799& 15.88&  0.08 \\
 S90 11238&none&   44931.5403& 15.62&  0.09 \\
 S90 11273&none&   44941.4799& 15.61&  0.11 \\
 S90 11328&GG13&   44988.4722& 16.11&  0.06 \\
 S90 11416&GG13&   45026.5333& 15.85&  0.05 \\
 S90 11462&GG13&   45053.4306& 15.91&  0.06 \\
 S90 11535&GG13&   45105.4257& 15.68&  0.07 \\
 S90 11847&none&   45343.4743& 16.01&  0.06 \\
 S90 11934&none&   45367.4444& 16.16&  0.06 \\
 S90 11981&none&   45398.4174& 16.12&  0.08 \\
 S90 12029&none&   45407.4458& 16.06&  0.05 \\
 S90 12057&none&   45440.3674& 16.24&  0.09 \\
 S90 12307&GG13&   45647.5139& 17.05&  0.05 \\
 S90 12453&GG13&   45766.5125& 16.88&  0.05 \\
 S90 12483&GG13&   45814.3646& 17.20&  0.05 \\
 S90 12669&GG13&   46027.5111& 16.67&  0.04 \\
 S90 12748&GG13&   46062.4549& 16.71&  0.09 \\
 S90 13784&GG13&   46909.4167& 16.77&  0.05 \\
 S90 13998&GG13&   47184.5174& 16.77&  0.07 \\
 S90 14019&GG13&   47205.4882& 16.56&  0.04 \\
 S90 14052&GG13&   47212.5042& 16.54&  0.04 \\
 S90 14066&GG13&   47231.3722& 16.62&  0.03 \\
 S90 14413&GG13&   47556.4083& 16.44&  0.04 \\
 S90 14427&GG13&   47558.4354& 16.40&  0.05 \\
 S90 14984&GG13&   48275.5528& 16.30&  0.09 \\
 OSC SJ03821& GG395& 48296.8333& 16.19&  0.07 \\
 S90 15014&GG13&   48301.4118& 16.34&  0.12 \\
 S90 15031&GG13&   48329.3840& 16.46&  0.05 \\
 S90 15136&GG13&   48628.5125& 15.98&  0.05 \\
SDSS ccd &  g &     52263.0000 &16.20 & 0.05 \\
VALLIN  ccd &   B  &    54995.3625 & 15.33 & 0.16 \\
\enddata
\end{deluxetable}

\begin{deluxetable}{rrrrr}
\tablecolumns{5}
\tablewidth{0pc}
\tablecaption{Historic $B_J$ magnitudes of BZBJ1148+5254 form the Asiago archive}
\tablehead{
\colhead{Plate} & \colhead{Filter}   & \colhead{JD}    & \colhead{B$_J$} &
\colhead{err}    
}
\startdata
 OSC O59 &none &33360.8611& 16.18& 0.10 \\
 S90 00201&none&   39181.4630& 16.32&  0.06 \\
 S90 00235&none&   39201.4450& 16.34&  0.08 \\
 S90 00484&none&   39477.5120& 16.45&  0.07 \\
 S90 00561&none&   39503.4980& 16.29&  0.06 \\
 S90 00625&GG13&   39565.6040& 16.40&  0.10 \\
 S90 00654&none&   39610.3470& 16.51&  0.06 \\
 S90 01482&GG13&   39938.3600& 16.43&  0.05 \\
 S90 02382&GG13&   40362.5150& 16.48&  0.08 \\
 S90 02389&GG13&   40364.4920& 16.38&  0.10 \\
 S90 02403&GG13&   40375.3970& 16.63&  0.11 \\
 S90 02404&GG13&   40386.3900& 16.53&  0.09 \\
 S90 02405&GG13&   40387.3900& 16.55&  0.10 \\
 S90 02420&GG13&   40406.4180& 16.88&  0.10 \\
 S90 04273&GG13&   41040.4990& 16.74&  0.06 \\
 S90 04301&GG13&   41057.4030& 16.79&  0.04 \\
 S90 05319&none&   41393.4840& 16.91&  0.12 \\
 S90 07151&GG13&   42134.3530& 16.76&  0.08 \\
 S90 07155&GG13&   42149.4560& 16.82&  0.07 \\
 S90 07870&GG13&   42453.5000& 16.91&  0.05 \\
 S90 07893&GG13&   42511.5220& 16.82&  0.04 \\
 S90 07898&GG13&   42514.4610& 16.85&  0.05 \\
 S90 08416&GG13&   42835.4810& 16.96&  0.03 \\
 S90 08532&GG13&   42889.4340& 16.73&  0.10 \\
 S90 08854&GG13&   43140.5700& 16.99&  0.04 \\
 S90 08928&GG13&   43191.5140& 16.91&  0.05 \\
 S90 08951&GG13&   43226.4930& 16.89&  0.05 \\
 S90 09444&GG13&   43311.3970& 16.73&  0.11 \\
 S90 09472&GG13&   43553.5480& 16.78&  0.09 \\
 S90 09486&GG13&   43631.3660& 16.95&  0.05 \\
 S90 09838&GG13&   43846.5620& 16.75&  0.09 \\
 S90 09910&GG13&   43905.5020& 16.81&  0.04 \\
 S90 09918&GG13&   43926.4350& 17.04&  0.08 \\
 S90 09943&GG13&   43936.4310& 16.96&  0.05 \\
 S90 09961&GG13&   43960.3160& 16.83&  0.06 \\
 S90 09967&GG13&   43962.3930& 16.79&  0.06 \\
 S90 09969&GG13&   43963.3750& 16.85&  0.05 \\
 S90 09984&GG13&   43966.4140& 16.95&  0.11 \\
 S90 10000&GG13&   43986.4000& 16.89&  0.07 \\
 S90 10028&GG13&   44013.4060& 16.86&  0.10 \\
 S90 10061&GG13&   44017.4480& 16.76&  0.09 \\
 S90 10309&GG13&   44219.4630& 16.92&  0.07 \\
 S90 10372&none&   44252.3540& 17.13&  0.18 \\
 S90 10391&none&   44275.3410& 17.09&  0.11 \\
 S90 10817&none&   44630.3770& 17.03&  0.10 \\
 S90 10887&none&   44640.4960& 16.79&  0.09 \\
 S90 11240&none&   44931.6040& 16.87&  0.05 \\
 S90 11305&GG13&   44972.6130& 16.88&  0.06 \\
 S90 11329&GG13&   44988.4990& 16.99&  0.05 \\
 S90 11452&GG13&   45048.4390& 17.02&  0.10 \\
 S90 11481&GG13&   45055.4430& 16.90&  0.04 \\
 S90 11502&GG13&   45086.4250& 17.00&  0.06 \\
 S90 11855&none&   45344.4910& 17.05&  0.07 \\
 S90 11942&none&   45378.4680& 16.92&  0.05 \\
 S90 12028&none&   45407.4260& 17.07&  0.06 \\
 S90 12059&none&   45440.4140& 16.97&  0.09 \\
 S90 12455&GG13&   45766.5700& 17.10&  0.08 \\
 S90 12482&GG13&   45813.4320& 16.87&  0.11 \\
 S90 12485&GG13&   45814.4090& 17.02&  0.10 \\
 S90 12700&none&   46035.5630& 17.00&  0.07 \\
 S90 12711&GG13&   46049.3910& 16.77&  0.12 \\
 S90 12746&GG13&   46061.4730& 17.02&  0.06 \\
 S90 12829&GG13&   46171.3960& 17.09&  0.10 \\
 S90 13211&none&   46419.5460& 17.12&  0.09 \\
 S90 13231&GG13&   46442.5120& 17.16&  0.06 \\
 S90 13292&GG13&   46478.4550& 17.40&  0.07 \\
 S90 13303&GG13&   46495.4500& 17.18&  0.10 \\
 S90 13335&none&   46563.3790& 17.06&  0.08 \\
 S90 13602&GG13&   46770.5420& 17.09&  0.04 \\
 S90 13756&GG13&   46879.4530& 17.01&  0.05 \\
 S90 13779&GG13&   46908.3580& 17.07&  0.04 \\
 S90 13796&none&   46937.3680& 17.23&  0.12 \\
 S90 14025&GG13&   47208.3990& 17.24&  0.04 \\
 S90 14069&GG13&   47231.4520& 17.02&  0.03 \\
 S90 14422&GG13&   47560.5290& 17.17&  0.03 \\
 S90 14487&GG13&   47586.4520& 17.05&  0.05 \\
 S90 14516&GG13&   47593.3880& 17.19&  0.05 \\
 S90 14535&GG13&   47651.3990& 17.11&  0.07 \\
 S90 15374&GG13&   48987.4730& 17.31&  0.10 \\
 OSC SJ05730& GG395 &49456.7896&17.20& 0.05\\
 SDSS ccd& b  &                        52288.0000&17.17& 0.03 \\
 VALLIN ccd  & B   & 54995.3500 &16.61& 0.09 \\
\enddata
\end{deluxetable}

\begin{deluxetable}{rrrrr}
\tablecolumns{5}
\tablewidth{0pc}
\tablecaption{Historic $B_J$ magnitudes of BZBJ1209+4119  form the Asiago archive}
\tablehead{
\colhead{Plate} & \colhead{Filter}   & \colhead{JD}    & \colhead{B$_J$} &
\colhead{err}    
}
\startdata
 OSC O1367 & none& 35190.8611& 17.50 & 0.10\\  
 S90 00184 &none&   39171.5993& 18.10&  0.08 \\
 S90 00237 &none&   39201.4681& 18.15&  0.13 \\
 S90 00297 &none&   39287.4389& 17.78&  0.17 \\
 S90 00631 &none&   39585.4861& 18.00&  0.09 \\
 S90 00648 &none&   39593.3966& 18.15&  0.12 \\
 S90 00692 &none&   39622.4952& 17.76&  0.13 \\
 S90 01519 &GG13&   39942.5063& 17.54&  0.08 \\
 S90 05377 &none&   41449.3452& 17.54&  0.08 \\
 S90 06896 &GG13&   42016.6196& 17.62&  0.21 \\
 S90 07667 &none&   42391.5167& 17.73&  0.05 \\
 S90 07748 &GG13&   42402.5980& 17.77&  0.10 \\
 S90 07851 &GG13&   42451.5091& 18.10&  0.06 \\
 S90 07905 &GG13&   42515.4299& 18.02&  0.07 \\
 S90 07942 &GG13&   42541.4966& 17.66&  0.07 \\
 S90 07976 &GG13&   42569.3959& 17.80&  0.07 \\
 S90 08368 &GG13&   42817.5660& 17.87&  0.04 \\
 S90 08461 &GG13&   42866.4771& 17.70&  0.08 \\
 S90 08523 &GG13&   42872.5396& 17.49&  0.08 \\
 S90 08530 &GG13&   42889.3917& 17.61&  0.13 \\
 S90 08567 &GG13&   42904.5466& 17.81&  0.12 \\
 S90 08907 &GG13&   43165.4924& 17.65&  0.13 \\
 S90 08922 &GG13&   43189.5119& 17.34&  0.05 \\
 S90 08954 &GG13&   43226.5675& 17.68&  0.07 \\
 S90 09013 &GG13&   43284.4786& 17.41&  0.08 \\
 S90 09034 &GG13&   43307.4015& 17.60&  0.15 \\
 S90 09499 &GG13&   43657.4376& 18.00&  0.11 \\
 S90 09928 &GG13&   43927.5189& 18.21&  0.04 \\
 S90 10052 &GG13&   44016.4473& 17.86&  0.12 \\
 S90 10346 &none&   44226.5035& 17.98&  0.13 \\
 S90 10906 &none&   44642.5404& 17.54&  0.09 \\
 S90 10947 &none&   44691.4446& 17.42&  0.14 \\
 S90 11337 &GG13&   44989.4911& 18.61&  0.08 \\
 S90 11453 &GG13&   45048.4675& 18.15&  0.09 \\
 S90 11482 &GG13&   45055.4723& 18.36&  0.09 \\
 S90 11518 &GG13&   45088.5147& 18.48&  0.10 \\
 S90 11877 &none&   45347.4869& 18.69&  0.08 \\
 S90 11885 &none&   45350.4841& 18.87&  0.12 \\
 S90 11989 &none&   45399.4445& 17.98&  0.05 \\
 S90 12040 &none&   45412.4348& 17.65&  0.10 \\
 S90 12055 &none&   45439.5035& 17.56&  0.09 \\
 S90 12441 &GG13&   45738.4689& 17.34&  0.11 \\
 S90 12493 &GG13&   45816.5279& 17.93&  0.10 \\
 S90 14075 &GG13&   47232.4258& 17.14&  0.07 \\
 S90 14382 &GG13&   47535.5251& 17.05&  0.09 \\
 S90 14442 &GG13&   47564.4377& 17.34&  0.14 \\
 S90 14994 &GG13&   48277.5331& 17.14&  0.15 \\
 S90 15022 &GG13&   48306.5585& 16.88&  0.09 \\
 S90 15043 &GG13&   48331.4518& 17.27&  0.11 \\
 OSC SJ06694&GG395&  50161.9111& 17.50 & 0.10\\
 SDSS ccd& b                    & 52731.0000 & 18.30& 0.05\\
 VALLIN ccd & B                & 55040.3560 & 18.30 & 0.20\\
\enddata
\end{deluxetable}

\clearpage
\begin{figure}
\includegraphics[scale=1.0]{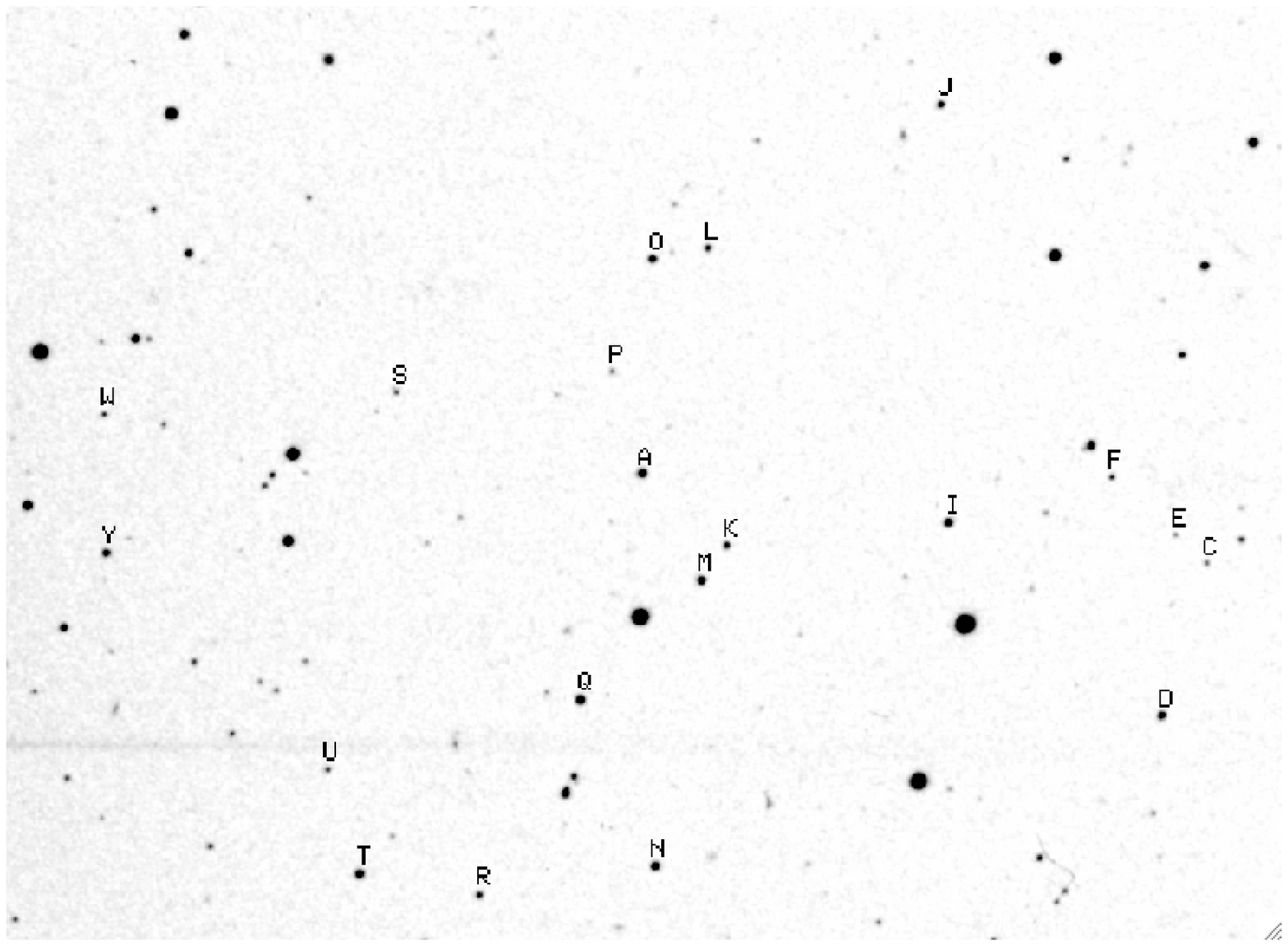}
\caption{Comparison stars in the field of BZBJ1058+56:
the reference stars are marked with letters starting from B,  the AGN is marked 
with A. North is up and East to the left. Field of view about 16' in DEC.
\label{cartina1058}
}
\end{figure}
\clearpage
\begin{figure}
\includegraphics[scale=1.00]{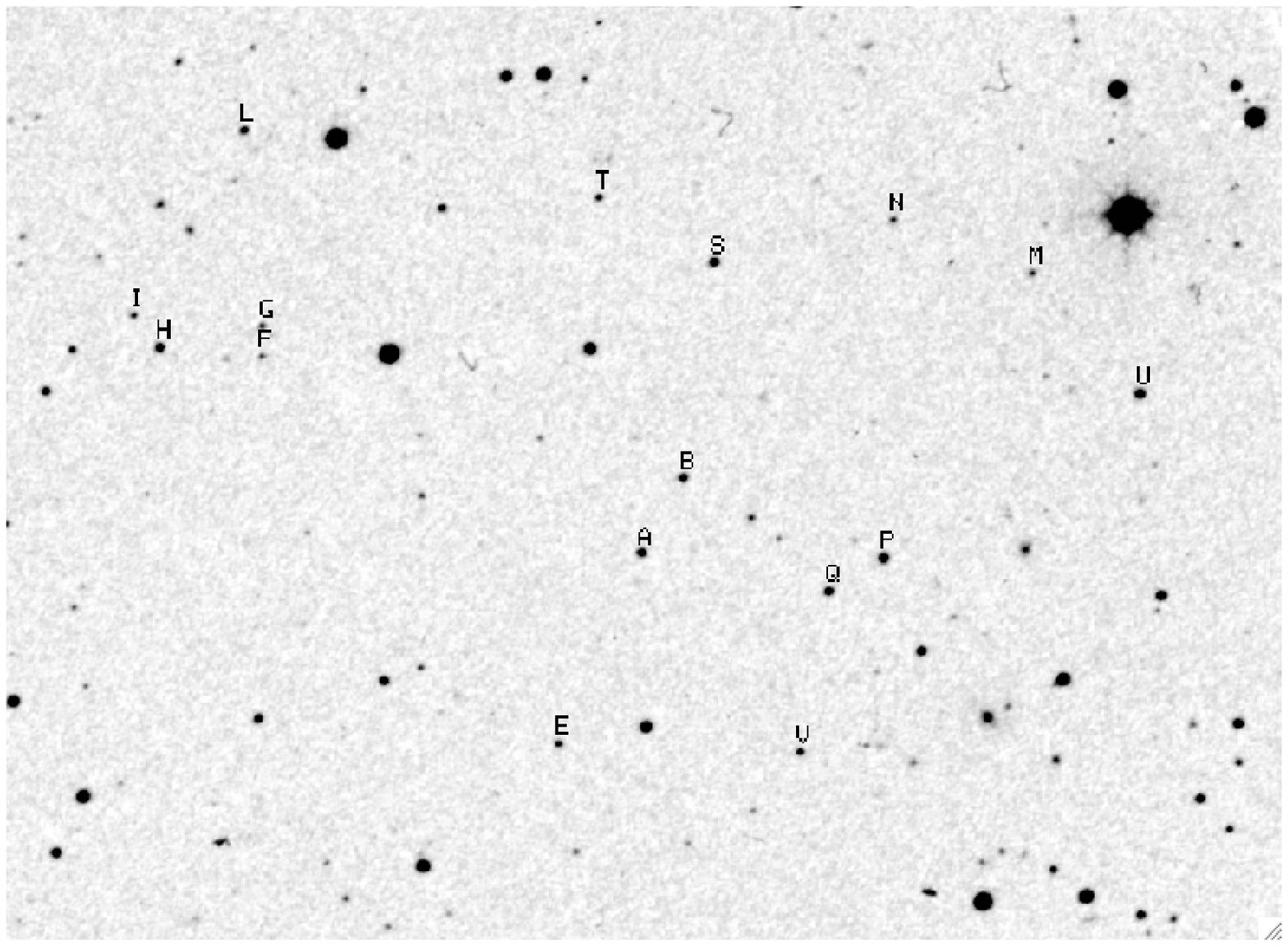}
\caption{Comparison stars in the field of BZBJ1148+52:
the reference stars are marked with letters starting from B,  the AGN is marked 
with A. North is up and East to the left. Field of view about 17' in DEC.
\label{cartina1148}
}
\end{figure}

\clearpage
\begin{figure}
\includegraphics[scale=1.00]{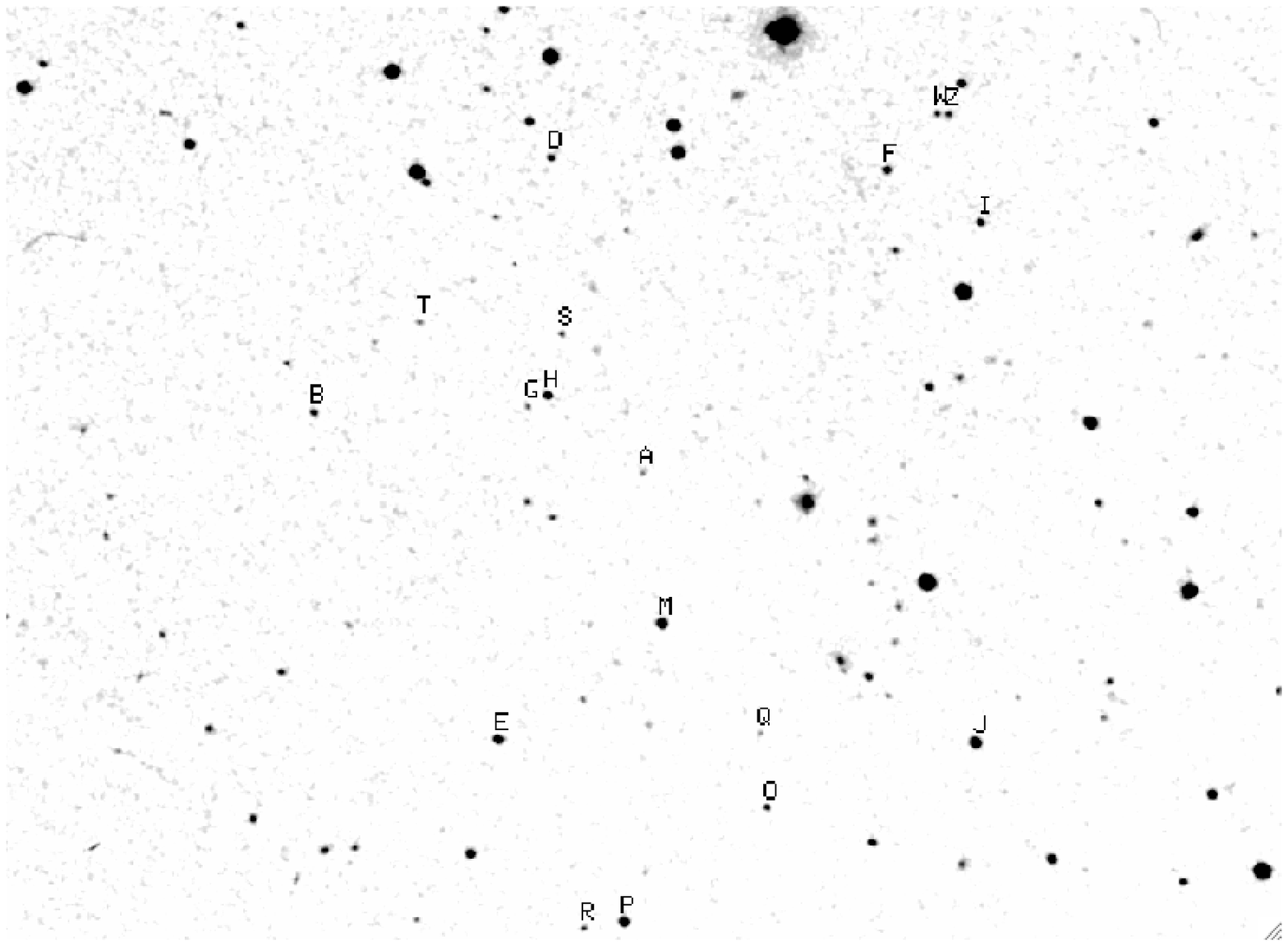}
\caption{Comparison stars in the field of BZBJ1209+41:
the reference stars are marked with letters starting from B,  the AGN is marked 
with A. North is up and East to the left. Field of view about 18' in DEC.
\label{cartina1209}
}
\end{figure}
\begin{figure}
\includegraphics[scale=1.0]{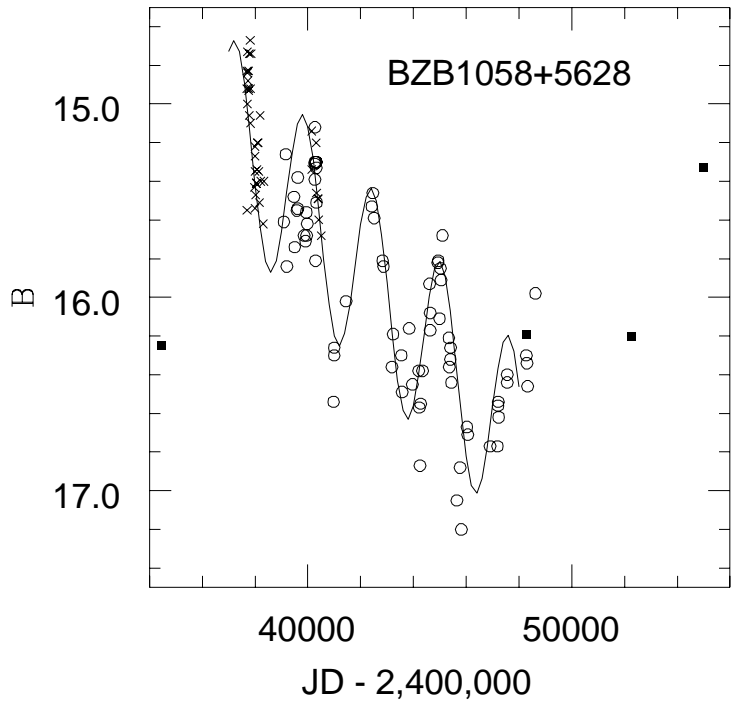}
\caption{The historic light curve in the $B$ band of BZBJ1058+56. 
from March 1953 to June 2009. 
Filled circles are data from the Asiago 67/92cm Schmidt plates; crosses from the Asiago 40/50cm Schmidt corrected for the zero point shift; open squares other data. Error bars are omitted for clarity. A sinusoidal line with a period of 2590 days and a monotonic decreasing trend is reported for comparison.
\label{LC1058}
}
\end{figure}
\begin{figure}
\includegraphics[scale=1.00]{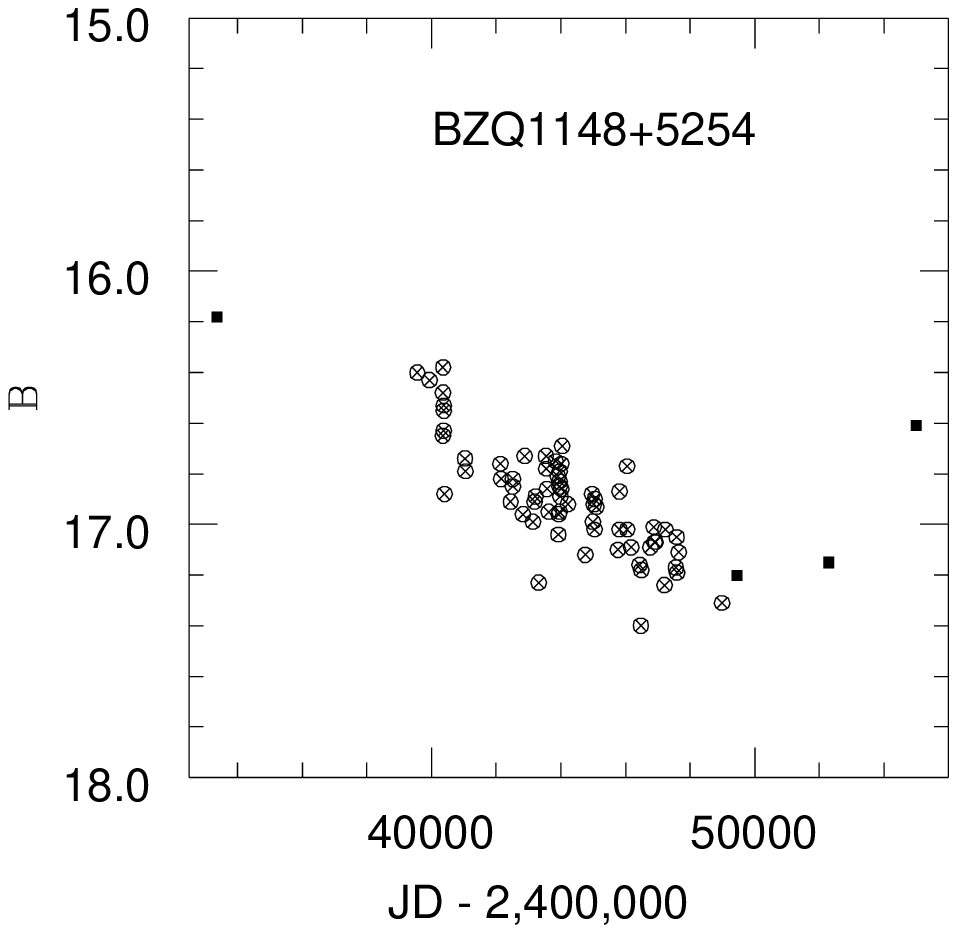}
\caption{The historic light curve in the $B$ band of BZBJ1148+52. 
from March 1950 to June 2009. 
Filled circles are data from the Asiago 67/92cm Schmidt plates with GG13 filter; crosses are unfiltered data from the same telescope corrected for the zero point shift; open squares are other data. Error bars are omitted for clarity. 
\label{LC1148}
}
\end{figure}

\begin{figure}
\includegraphics[scale=1.00]{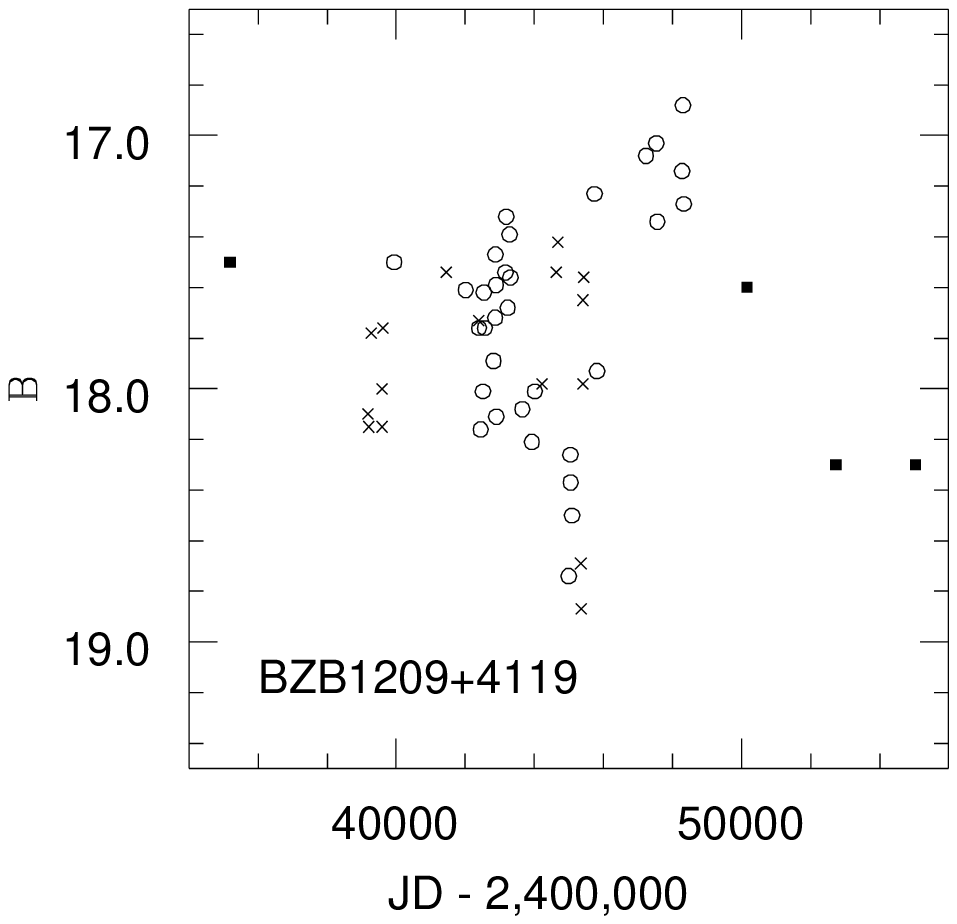}
\caption{The historic light curve in the $B$ band of BZBJ1209+41. 
from March 1955 to July 2009. 
Filled circles are data from the Asiago 67/92cm Schmidt plates with GG13 filter;
crosses are unfiltered data from the same telescope corrected for the zero point shift; open squares are other data. Error bars are omitted for clarity. 
\label{LC1209}
}
\end{figure}

\begin{figure}
\includegraphics[scale=1.00]{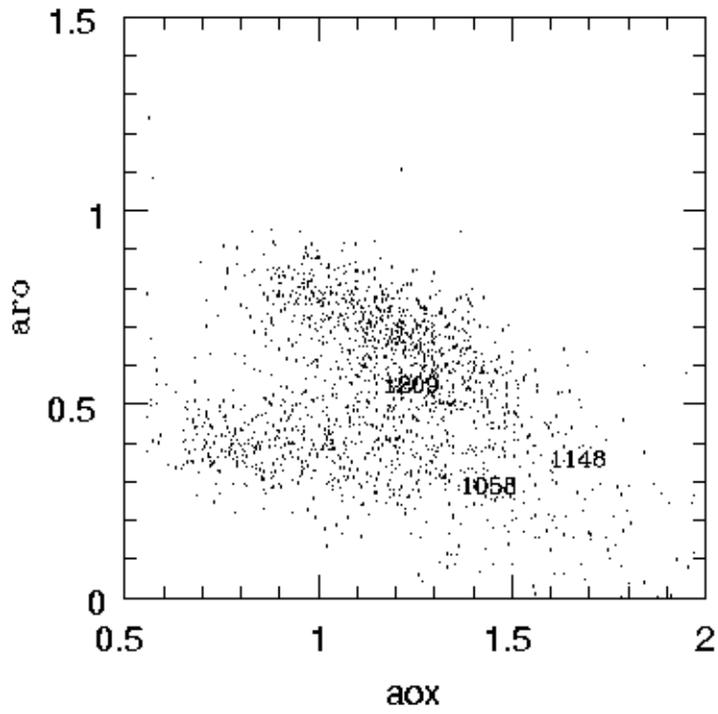}
\caption{The $\alpha_{ro} - \alpha_{ox}$ plot for the sources in the Roma BZCat. The approximate positions of the three sources of this paper are marked.
\label{aroaox}
}
\end{figure}

\end{document}